\documentclass[usenatbib,useAMS,preprint,psfig2]{mn2e}
\usepackage{times}
\usepackage{amsmath}
\usepackage{amssymb,amsfonts}
\usepackage{graphicx}
\usepackage{epstopdf}
\usepackage{verbatim}
\input psfig.sty

\def\be{\begin{equation}}
\def\ee{\end{equation}}
\def\bea{\begin{eqnarray}}
\def\eea{\end{eqnarray}}

\def\nn{\nonumber}
\def\mcr{{{\rm M_{cr}}}}
\def\xo{{x_{o}}}
\def\dm{\Delta {\rm M}}
\def\ms{{\rm M_{\odot}}}
\def\mb{m_{B}}
\def\bo{B_{o}}

\def\rsix{R_{6}}
\def\vr{v_{{\rm r}}}
\def\vro{v_{{\rm ro}}}
\def\vt{v_{\theta}}

\def\ov{ \over}

\def\rmo{{\rm R_{M0}}}
\def\rrm{{R_{{\rm M}}}}
\def\rra{{R_{{\rm A}}}}

\def\mdot{\ifmmode \dot M \else $\dot M$\fi}    
\def\mxd{\ifmmode \dot {M}_{x} \else $\dot {M}_{x}$\fi}
\def\med{\ifmmode \dot {M}_{Edd} \else $\dot {M}_{Edd}$\fi}
\def\bff{\ifmmode B_{{\rm f}} \else $B_{{\rm f}}$\fi}

\def\apj{\ifmmode ApJ \else ApJ \fi}    
\def\apjl{\ifmmode  ApJ \else ApJ \fi}    %
\def\aap{\ifmmode A\&A \else A\&A\fi}    %
\def\mnras{\ifmmode MNRAS \else MNRAS \fi}    %
\def\nat{\ifmmode Nature \else Nature \fi}
\def\prl{\ifmmode Phys. Rev. Lett. \else Phys. Rev. Lett.\fi}
\def\prd{\ifmmode Phys. Rev. D. \else Phys. Rev. D.\fi}

\def\ms{\ifmmode M_{\odot} \else $M_{\odot}$\fi}    
\def\na{\ifmmode \nu_{A} \else $\nu_{A}$\fi}    
\def\nk{\ifmmode \nu_{K} \else $\nu_{K}$\fi}    
\def\ns{\ifmmode \nu_{{\rm s}} \else $\nu_{{\rm s}}$\fi}
\def\no{\ifmmode \nu_{1} \else $\nu_{1}$\fi}    
\def\nt{\ifmmode \nu_{2} \else $\nu_{2}$\fi}    
\def\ntk{\ifmmode \nu_{2k} \else $\nu_{2k}$\fi}    
\def\dnmax{\ifmmode \Delta \nu_{max} \else $\Delta \nu_{2max}$\fi}
\def\ntmax{\ifmmode \nu_{2max} \else $\nu_{2max}$\fi}    
\def\nomax{\ifmmode \nu_{1max} \else $\nu_{1max}$\fi}    
\def\nh{\ifmmode \nu_{\rm HBO} \else $\nu_{\rm HBO}$\fi}    
\def\nqpo{\ifmmode \nu_{QPO} \else $\nu_{QPO}$\fi}    
\def\nz{\ifmmode \nu_{o} \else $\nu_{o}$\fi}    
\def\nht{\ifmmode \nu_{H2} \else $\nu_{H2}$\fi}    
\def\ns{\ifmmode \nu_{s} \else $\nu_{s}$\fi}    
\def\nb{\ifmmode \nu_{{\rm burst}} \else $\nu_{{\rm burst}}$\fi}
\def\nkm{\ifmmode \nu_{km} \else $\nu_{km}$\fi}    
\def\ka{\ifmmode \kappa \else \kappa\fi}    
\def\dn{\ifmmode \Delta\nu \else \Delta\nu\fi}    

\title{The bottom magnetic field and magnetosphere evolution of
neutron star  in low mass X-ray binary}
\author[C.M. Zhang and Y. Kojima]{C.M. Zhang$^{1}$  and Y. Kojima$^{2}$\\
1.  National Astronomical Observatories,
 Chinese Academy of Sciences, Beijing 100012, China \thanks{E-mail: zhangcm@bao.ac.cn}\\
2.  Department of Physics, Hiroshima University, Higashi-Hiroshima
739-8526, Japan}

\begin{document}

\date{Received date ; accepted date}

\maketitle

\begin{abstract}
 The accretion induced neutron star magnetic field evolution is
studied through considering the accretion flow to drag the field
lines aside and dilute the  polar field strength, and as a result
 the equatorial field strength increases, which  is buried inside
the crust on account of the accretion induced global compression
of star crust.  The  main conclusions of model are as follows:
  (i) the polar  field decays with increasing the accreted mass;
(ii) The bottom magnetic field strength of about $10^8$ G  can
occur when neutron star magnetosphere radius approaches  the star
radius,  and it  depends on  the  accretion rate as $\mdot^{1/2}$;
 (iii) The neutron star magnetosphere radius decreases with accretion
until it reaches the star radius, and  its evolution is little
influenced by the initial field and the accretion rate
 after  accreting $\sim 0.01 \ms$, which implies that
 the magnetosphere radii  of neutron stars in LMXBs
 would be homogeneous  if they accreted the comparable masses.
    As an extension, the physical  effects of
 the possible strong magnetic zone in the X-ray neutron stars
 and recycled pulsars are discussed.
 Moreover,  the strong magnetic fields  in
the binary pulsars  PSR 1831-00 and
PSR 1718-19 after accreting  about half solar mass in
the binary accretion phase,
$8.7\times10^{10}$ G and $1.28\times10^{12}$ G,
 respectively,
can be explained  through  considering the incomplete frozen flow
in the polar zone. As a model's expectation,
 the existence of the low magnetic field
($\sim 3\times 10^{7}$ G) neutron stars or millisecond pulsars is
suggested.
\end{abstract}




\section{Introduction}
 Recently, the newly discovered double pulsars
J0737-3039A and  J0737-3039B  in  the binary system
(Burgay et al 2004; Lyne et al 2004; van den Heuvel 2004; Lorimer 2004)
have  shown that the millisecond pulsar (MSP) possesses
 the low field  and short   spin period
($B=7\times10^{9}$ G, P=22.7 milliseconds) and the normal pulsar
possesses high field and long spin period ($B=6\times10^{12}$ G,
P=2.77 seconds), which convinces our previous  point of view that
 MSP is recycled in the binary system where the accreted  matter
weakened its magnetic field
 and accelerated its  spin (see, e.g., van den Heuvel  2004).
 Moreover, with the launch of  RXTE satellite,  the
 accretion-powered X-ray pulsar SAX J 1808.4-3658
  (P=2.49 milliseconds),
as well as  other five examples (see, e.g., Chakrabarty 2004; van
der Klis 2004; Wijnands \& van der Klis 1998), and 11 type-I X-ray
burst frequency,  confirmed as stellar  spin frequency, sources have
been discovered (see, e.g., Muno 2004; Chakrabarty 2004; van der
Klis 2004, 2000), which exhibits the strong evidence that neutron star
(NS) in low mass X-ray binary (LMXB) is the progenitor of MSP with
spin frequency $\sim $ 400 Hz and magnetic field strength $\sim
10^{8-9}$ G because of the accretion
 (see, e.g., Wijnands \& van der Klis  1998; Chakrabarty 2004).

The origin, structure and evolution of NS  magnetic field is still
an open problem (see, e.g., Bhattacharya \& van den Heuvel 1991;
Phinney \& Kulkarni 1994; Bhattacharya \& Srinivasen 1995; Colpi
et al
 2001; Melatos \& Phinney 2001; Cumming 2004;
 Blandford et al  1983;  Blondin \& Freese 1986).
   In the early stage of discovery of NS, it is believed that
 NS magnetic field decays  due to Ohmic
dissipation  in the crust. But later calculations of Ohmic
dissipation
suggest that isolated NS magnetic field  may not decay
significantly (Sang \& Chanmugam  1987), if the field occupies
 the entire
crust, and remains large for  more than  Hubble time.
 Although many instructive
 mechanisms on the isolated NS magnetic decay  have been proposed,
 such as,  the crustal plate tectonics model by Ruderman (1991)
 and Chen \& Ruderman 1993,
 the Hall-drift dominated field evolution
  (see, e.g., Rheinhardt et al 2004; Jones 2004;
  Hollerbach \& Ruediger 2002;  Geppert \& Rheinhardt 2002;
 Rheinhardt \& Geppert 2002; Naito \& Kojima 1994), the
 spin-evolution  induced magnetic field decay (see, e.g.,
 Konar \& Bhattacharya 1997, 1999;
 Ding et al 1993; Jahan Miri \& Bhattacharya 1994; Ruderman et al 1998),
 etc., there has not yet been a commonly accepted idea on such issue.

 The accretion induced NS magnetic field decay  has not
been paid much attention  until 1980's when Taam and van den
Heuvel (1986) presented the NS magnetic field evolution associated
with the accretion  in the binary system, based on which Shibazaki
et al (1989) concluded the simple empirical  formula of the field
deduction versus the accreted mass. However, it is indicated that
the significant decay of magnetic field is achieved only if the
neutron star experiences the interacting binary (see, e.g.,
Verbunt \& van den Heuvel 1995; van den Heuvel 1995 and references
therein).  Furthermore,  van den Heuvel and Bitzaraki (1995),
 from the statistical analysis of 24
binary radio pulsars with nearly circular orbits and low mass
companions, discovered a clear correlation between spin period and
orbital period, as well as the magnetic field and orbital period.
These relations strongly suggest that the increased amount of
 accreted mass  leads to the decay of NS  polar magnetic field, and the
`bottom' field strength  $\sim 10^{8}$ G is also implied, which
means that NS magnetic field evolution will stop at some minimum  value,
 $\sim10^{8}$ G
 (see also, e.g., Burderi et al 1996; Burderi \& D'Amico 1997).


 In order to understand
  the accretion induced field decay, some suggestions and
models have been proposed. Based on the accretion buried or
screened NS magnetic field guess by  Bisnovati-Kogan and Komberg
(1974),
  Zhang et al (1994, 2000) proposed
  the ferromagnetic crust screen model   to interpret
  the simple
inverse correlation between the field deduction  and the accreted
mass, as  declaimed by Shibazaki et al (1989).
  Moreover, the accretion induced flow
and thermal effects to account for the speed-up Ohmic dissipation
of NS crust currents are also proposed  and studied by several
authors
 (see, e.g., Romani 1990;  Geppert \& Urpin 1994; Urpin \& Geppert 1995).
  Later on, the diamagnetic screening of NS magnetic field by freshly
accreted material is suggested (see e.g. Lovelace et al. 2005),
 which would be effective above a
critical accretion rate $\mdot$ of a few percent of the Eddington
rate (see, e.g., Cumming et al 2001).
 More recently, it is calculated that
 the polar cap widening or
   the equator-ward hydro-magnetic spreading   by the accreted
 material flowing accounts for
the NS magnetic field decay (see, e.g., Konar \& Choudhury 2004,
references therein; Payne 2005; Payne \&  Melatos 2004;  Melatos \& Phinney
2001).  Furthermore, the magnetic burial mechanism has been
studied in detail analytically and numerically  by Payne \&  Melatos
(2004) and Payne (2005) under the assumptions of  no instabilities or Ohmic
diffusion, where the structure of highly distorted magnetic field
and hence the magnetic  moment as a function of accreted mass is
computed. They find  that   the deduction of the magnetic moment is
inversely related to how much mass accreted and about $\sim 10^{-5}$\ms
is required  to reduce significantly the magnetic dipole moment.


In this paper,
 we follow   the idea,  firstly proposed  by van den
Heuvel and Bitzaraki (1995),  and also by Burderi and D'Amico
(1997),  that    the accretion matter is channeled by the strong
magnetic field initially into the two polar patches, corresponding
to NS
 magnetosphere radius of several hundreds of stellar radius, pushes the
field lines aside and thus dilutes the polar field strength due to
the flow of  accreted  materials from the polar to the equator.
The bottom field should be reached when the accretion becomes all
over star, which corresponds to that the NS
 magnetosphere radius matches the star radius,
  and in turn it  infers the bottom  magnetic field
   to be $\sim 10^{8}$ G.

The paper is organized  as follows: in the next section the model
is described and the magnetic field evolution  equation  is
derived. Section 3  outlines the applications of the model,
including how the NS magnetic field and magnetosphere evolve with
the accreted mass, the initial magnetic field and accretion rate,
etc.. The discussions  are drawn in Section 4, where we summarize
the conclusions of model and the  observational consequences.

\section{ Description and calculation of the model}

\subsection{Dilution of the polar magnetic flux}

For  the spherical  magnetic slab   geometrical  structure of
accreted NS  illustrated in Fig.1, we assume that the magnetic
field lines are  anchored, completely frozen or incompletely
frozen,   in the entire NS crust  with constant average mass
density.
 The  introduction of the spherical topological
magnetic polar cap geometry  is on account of its important
application in describing  the
 kHz QPO mechanism (Zhang 2004),
where  the  spherical magnetic polar cap geometry in Eq.(\ref{ap})
is used  and the approximated  plane
 magnetic slab (Cheng \& Zhang 1998) is inadequate
 when the magnetosphere   is  close to  the star surface. 
Moreover, in the work by Cheng \& Zhang (1998) the 
rate of change of the magnetic polar cap area   is roughly 
performed (see Eq.(5) of that paper), and it should be derived from 
the magnetic flux conservation as described below.  
  Initially,
the magnetic field is sufficient strong, and the accreted matter
will be channeled into the polar patches by the field lines. The
cumulated accreted matter will be compressed into the crust region
through the polar patch or through the surface flow on account of
the plasma instabilities (see, e.g., Litwin et al.  2001). 
However, as a preliminary exploration, we  neglect these instability
effects on the polar field deduction, and concentrate on the
description of the field evolution by the assumed frozen-in crust
MHD motions. 
We suppose the compressed accreted matter to arise the expansion
of the magnetic polar zone into two directions, downward and
equator-ward.
  Under the condition of  incompressible fluid approximation and  the
constant crustal volume assumption, as well as  the averaged
homogenous NS mass density  $\rho \sim 10^{14} g/cm^{3}$,  where
the Ohmic dissipation time scale of field is longer than the flow
time scale for  sub-Eddington   accretion  rates (see, e.g., Brown \&
Bildsten 1998; Geppert et al 1999),
 the magnetic line  frozen motion is assumed
 and the incomplete frozen motion may appear if the  unknown
  instabilities are taken into effects (see, e.g., Melrose 1986).
Therefore, in a particular duration
$\delta t$ of accretion with the mass accreting rate $\mdot$,
the piled accreted mass $\delta M = \mdot \delta t$
  will arise the expansion of the volume of the  magnetic
polar zone  in  the crust, by virtue of the mass flux conservation,

\be
(\delta M/2 - \rho A_{p} \delta H ) \xi = \rho  H  \delta A_{p}\,,
\label{dvp} \ee where $A_p$ is the surface area of the magnetic
polar zone defined in Eq.(\ref{ap}).
 The introduced parameter $\xi$ ($0 \le\xi\le
1$) is an  efficiency parameter to express the incomplete magnetic
line  frozen flow by the plasma instability.  In two extremely
cases,  $\xi=1$ represents that the completely frozen  field lines
totally drift with the moving mass elements,  and  $\xi=0$
represents that the completely leaking of plasma makes  the field lines
 not drift with the moving mass elements.
 The physical meaning of Eq.(\ref{dvp}) is clear, $\xi=1$ for instance,
  that the r.h.s of Eq.(\ref{dvp}) is the mass flux toward the equatorial
  direction and the l.h.s of   Eq.(\ref{dvp}) is the mass accreted
  in one polar cap
  ($\delta M/2$) minus the mass sinked into the NS core.
H is the thickness of the crust and $\delta H$/H  is the fraction
of the thickness dissolving into the core,  defined by

\be \frac{\delta H}{H} =
\frac{\delta M}{M_{cr}} = \frac{\dot{M} \delta t}{M_{cr}}\,,
\label{dhh} \ee where $M_{cr}=4\pi R^2 \rho H$ is the crust mass.
For the standard neutron star structure model, the NS crust mass
is only a small portion of the entire NS mass, which is also
dependent of the equation of state (EOS)
 of matter inside NS (see, e.g., Lattimer \&
Prakash 2004; Baym \& Pethick 1975), so $M_{cr}/M \simeq 3H/R\sim
10\%$ is approximately implied. The expansion of the polar zone
will dilute the magnetic flux density if the magnetic flux
conservation is preserved, that is,

\be
\delta(BA_p)=0 \, \, \, \,  ,  \, \, \,
  \delta A_p = -A_p \frac{\delta B}{B}\,,
\label{dvp2}
\ee

and the area $A_p$ of the accretion polar
patch can be accurately expressed
by the angle  $\theta$, which is subtended at the stellar
surface    between the last
closed field line of the NS magnetosphere
  and the  polar axis,
\be
A_p = \int_{0}^{\theta}2\pi R^2 \sin\theta' d\theta' = 2\pi R^{2} (1-\cos\theta) \;.
\label{ap}
\ee
The polar field line can be  described by the
polar angle $\theta_{r}$  and the position r that satisfies
$(\sin^{2}\theta_{r})/r$=constant
(see, e.g. Shapiro \&  Teukolsky 1983, p453),
therefore, at the stellar surface, in terms of
the NS  magnetosphere radius $\rrm$,
 \be
sin^2\theta =\frac{R}{R_M}\,.
\label{sint}
\ee
If $\theta$ is small as a case of large magnetosphere radius,
 we have the approximated relation
$A_{p}=\pi R^{2}\sin^{2}\theta = \pi R^{3}/\rrm$, as conventionally used
(see, e.g. Shapiro \&  Teukolsky 1983,  p453).

$\rrm$ is related to the Alfven  radius $\rra$ by
\be
R_M = \phi R_A\, ,
\ee
\be
R_A =3.2\times 10^8 ({\rm cm})  \dot{M}^{-2/7}_{17} \mu^{4/7}_{30}
m^{-1/7}\,\,\, ,
\label{ra}
\ee
where $\phi$ is a parameter of about 0.5 (see, e.g., Ghosh \&  Lamb 1979;
Shapiro \& Teukolsky 1983), but  is  model dependent
(see, e.g., Li \& Wang 1999).
$m = M/M_{\odot}$ is the NS mass M in units of solar mass,
and   $\dot{M}_{17}$  and $\mu _{30}$  are  the accretion rate in
units of $10^{17}$ g/s and the magnetic moment in units
of $10^{30} {~ } {\rm G cm^3}$, respectively.

\subsection{The buried  magnetic flux  in the equatorial zone}
 With the proceeding of the accretion, the magnetic flux in the
polar zone will be diluted and the magnetic flux in the equatorial
zone would be increased because of the global conservation of the
magnetic flux. However, the new constructed magnetosphere radius
is  still determined by the new constructed polar magnetic field
because the increased equatorial magnetic flux is compressed into
the star crust by the global radial accretion  flow.
  Therefore the equatorial strong magnetic field lines are curved and
 buried inside
crust  and  might  be not seen during the accretion phase
 with the sufficient accretion rate, but it may be seen
 in the post-accretion phase on account of the Ohmic diffusion.
 We will try to confirm the above claim
  through comparing  the magnetic flux drifted from the
polar zone to the equatorial zone and the magnetic flux
 compressed into the crust  in the
equatorial zone.

We substitute Eq.(\ref{dhh}) into  Eq.(\ref{dvp}), and then obtain

\be
(2\pi R^2 - A_{p}) {\xi \delta H \over H\delta t}
= { \delta A_{p}\over \delta t} \,.
\label{dvp3} \ee
After considering Eq.(\ref{ap})
and making  the convenient arrangement, we can
obtain the ratio between the field line drift velocities in
the latitude  direction and in the radial direction,

\be {\vt \ov \vr} = {\xi R \cos \theta \ov H \sin \theta}\; ,
\label{vtvr}\ee with  the radial velocity $\vr=\delta H/\delta
t=\mdot/4\pi R^{2}\rho$ and the latitude  velocity  $\vt=R\delta
\theta/\delta t$.
 
The polar magnetic flux variation
 rate expelled from the polar zone with the approximated
radius $R\sin\theta$ in the latitude  direction
is
\be \dot{\Phi}_{in}\simeq 2\pi RB\sin\theta \vt\;, \ee
 with  $\dot{\Phi}_{in}={d\Phi_{in}\ov dt}$,
  and the equatorial  magnetic flux rate compressed
  into the crust in the radial direction
is \be \dot{\Phi}_{out}\simeq 2\pi RB \vro\;, \ee with
$\dot{\Phi}_{out}={d\Phi_{out}\ov dt}$ and $\vro=\mdot/4\pi
R^{2}\rho_{o}$, where $\rho_{o}$ is a characteristic mass density
 at where
 the matter pressure controls the magnetic pressure. If the
non-relativistic neutron in the crust is taken into account, the
matter pressure is expressed to be
 $P_{M}=k\rho^{5/3}$ with k=$5.38\times10^{9}$ in c.g.s. units
(Shapiro \& Teukolsky 1983). And the magnetic pressure $P_{B}$ is
calculated by the possible maximum permitted magnetic field
strength
 $B_{eqmax}\sim 10^{15}$ Gauss during the  accretion compression,
 exceeding over which
 the magnetic energy will be released by the soft gamma ray burst
 triggered by the super-strong magnetic induced crust cracking
 (see, e.g., Thompson \& Duncan 1995),
therefore we have, \be
B_{eqmax}^{2}/8\pi = k\rho_{o}^{5/3}\;, \ee or
 \be \rho_{o}
\simeq 10^{11} (g/cm^{3}) (B_{eqmax}/10^{15})^{6/5}\;. \ee  The
ratio between the flux variations in the polar zone and in the
equatorial zone can be  given if the condition of Eq.(\ref{vtvr})
is considered,
 \bea {\dot{\Phi}_{in}\ov \dot{\Phi}_{out}}
  & = & {\vt {}\sin\theta  \ov \vro} = {\xi R
\rho_{o}\cos\theta \ov H \rho}\\ \nn &\sim& 0.01 \xi \cos\theta
(B_{eqmax}/10^{15})^{6/5} < 1\;, \label{fratio}\eea
 and we find that this ratio is less than unity for the possible parameters,
 which means that the magnetic flux expelled into the
equatorial zone through the latitude  direction
 is much less than that compressed into the crust through the
 radial direction. Therefore,   we can conclude that
 the magnetic field strength at the magnetosphere-disk boundary
  will be dominated by the field lines from the polar cap zone,
  or the magnetosphere radius $\rrm$
   during accretion will be determined by the polar field. The
effective magnetic moment of star decreases because of the motions
of field lines,  dragged firstly to the equator zone from the
polar zone through the latitude  motion and then compressed into
the crust through the radial motion.
 Physically,  the open field line in the polar zone will be
 influenced by the
latitude  flow and little influenced by the radial flow, however
the closed field line in the equatorial  zone will  be dominated
by the radial flow.  If there exists
 only the  radial flow and no latitude  flow,
 as a case of the magnetosphere reaching the star surface,
   the polar field will be  little changed because the net
 motion of field lines in the latitude  direction
 disappear.  Therefore, with the proceeding of the accretion,
 the polar field decays and the newly constructed magnetosphere
radius is still determined by the decayed polar field strength, or
in mathematical terminology the field in the Eqs.(\ref{ap}) and
(\ref{ra}) will be same as the polar field described in
Eq.(\ref{dvp2}).

Furthermore, we stress that the above
arguments for the field line motion are roughly valid for the
purpose of the phenomenological model, however the detail
descriptions of the accretion flow, together with the magnetic
induction equations,
 are needed in presenting the NS magnetic structure, which will be
 the subsequent work elsewhere.

\subsection{The  polar magnetic field evolution}
 Furthermore, substituting Eqs.(\ref{dhh}),
 (\ref{dvp2})and (\ref{ap}) into Eq.(\ref{dvp}), we
obtain the magnetic field evolutionary equation as follows,

\be
\frac{A_p\delta B}{(2\pi R^2 - A_p)B}
=\frac{-\xi \dot{M}\delta t}{M_{cr}}\,,
\label{apdb}
\ee
or,
\be
\frac{(\sqrt{R_M} - \sqrt{R_M - R})\delta B}
{\sqrt{R_M  - R}B}=\frac{-\xi\dot{M}\delta t}{M_{cr}}\,.
\label{apdb2}
\ee
Solving Eq.(\ref{apdb}) or Eq.(\ref{apdb2}) with the initial condition
B(t=0)=$\bo$, the analytic magnetic field evolution
solution is  obtained,

\be\label{bt}
B = \frac{\bff}{\{1 - [C/\exp(y)-1]^2\}^{7/4}}\,,
\ee
with  $y=\frac{2\xi\Delta M}{7M_{cr}}$, $\dm=\mdot t$,
 $C = 1+\sqrt{1-x_0^2} \sim 2$  and
$ \xo^{2} = (\frac{\bff}{\bo})^{4/7} = R/\rmo$, where $\rmo$ is
the initial magnetosphere radius  and $B_f$ is the bottom
magnetic field
defined by the NS magnetosphere radius matching the star  radius,
i.e., $\rrm (\bff) = R$, therefore, \be \bff  = 1.32 \times 10^8\,
(G)\, (\frac{\dot{M}}{\dot{M}_{Ed}})^{1/2} m^{1/4}R^{-5/4}_6
\phi^{-7/4}\,, \label{bmin} \ee where $\dot{M}_{Ed}$ is the
Eddington accretion rate and $R_6$ is the NS radius in units of
$10^6$ cm.
  For the reason of simplicity, the usual  NS parameter
values
are set in the later  calculations and discussions, such as, m=1.4,
$\rsix=1.5$, $\phi=0.5$ and  $\mcr=0.2 \ms$.

\section{The applications of the model}

\subsection{ Inverse correlation between the magnetic
field and the accreted  mass}

In the  complete magnetic frozen case $\xi=1$ (for the incomplete
frozen case, we just take the replacement $\mcr/\xi \rightarrow
\mcr$), the magnetic field evolution with the accretion in
 Eq.(\ref{bt}) can be approximately simplified in
the following form by the Taylor serial expansion
with the condition $y\ll 1$ or
$\dm \ll 3.5 \mcr $,  
\be
B = \frac{\bo}{(1 + \frac{4\dm}{7\mb})^{7/4}}\,,
\label{shib0}
\ee
where $\mb = 0.5 \xo^{2}\mcr = 0.5(R/\rmo)\mcr$.
For the usual value of bottom field
$\bff\sim 10^{8}$ G and initial field  $\bo\sim 10^{12}$ G, we
obtain $\mb \sim 10^{-3} \ms$.
 For the high mass X-ray binary (HMXB)
or the low mass X-ray binary (LMXB) in the early stage, a little
mass is accreted, $\Delta M \sim  10^{-5} M_\odot \ll \mb$,
 then Eq.(\ref{shib0}) gives,
\be\label{shib}
B = \frac{B_0}{1 + \frac{\dm}{m_B}}\,,
\ee
which  is just the same form as the empirical formula
of the accretion induced field decay proposed by
Shibazaki et al. (1989) when fitting
the observational  data given by  Taam and van den Heuvel (1986).
 Shibazaki et al (1989) find that the best fitted mass constant is
$m_{B}$ = $10^{-5}$ -- $ 10^{-3}$ \ms
when comparing the theoretical curve of the magnetic field versus
the spin period  to the observational  data.
Moreover, the similar  mass value
 of significantly reducing  the magnetic dipole
moment  is also obtained
by Payne \& Melatos (2004) and Payne (2005) in their
 self-consistent analytic and numerical
 calculations of the accretion buried NS
polar magnetic field.

 If $\mb \ll \dm \ll 3.5 \mcr$, then we have the following
approximation from Eq.(\ref{bt}), \be B\simeq 0.8\bff ({\dm \ov
\mcr})^{-1.75}\,. \label{bdm2} \ee Therefore, Eq.(\ref{bdm2})
implies that the influence of the initial  magnetic field on the
magnetic evolution disappears at this stage, and the NS magnetic
field is scaled by the bottom field $\bff$.
 Furthermore,  with increasing of the accreted  mass
$\dm \rightarrow  \mcr$, the NS magnetic
evolution will enter into the "bottom state",  and
$B\rightarrow\bff$ is obtained.
As a detail illustration, the evolution of NS magnetic  field
versus the accreted  mass  with various  parameter conditions
 is
plotted in Fig.2, where we find that
  the influence by the  initial field  on the field
 evolution exists
when $\dm \ll 0.01 \ms$ and disappears when $\dm>0.01 \ms$.
Finally,
 the magnetic field evolution will enter into the "bottom  state" when
$\dm \rightarrow \mcr \sim 0.2 \ms$.

\subsection{ The bottom magnetic field
 and its correlation to the X-ray
luminosity}

The  conception of bottom  field, the possibly arrived minimum
field of accreted NS, is evidently proposed by van den Heuvel and
Bitzaraki (1995) from the analysis of the magnetic fields of the
millisecond pulsars in the binary systems versus the estimated
accreted masses,
which is  explained that the accreted matter pushes and dilutes
the polar field lines if the accretion in the channel way and
there is no net flow drag effect on the field lines if the
spherical accretion in the random way all over star begins.
 From the point of view of the accretion geometry,
 the bottom field is reached when the
accretion becomes isotropic all over star or
 the channeled accretion disappears completely
  (see also, e.g., Burderi et al 1996; Burderi \& D'Amico 1997).

 Mathematically, as a minimum field,
 the bottom magnetic field can also be determined by the condition
of vanishing the field variation respect to the accreted mass in
Eq.(\ref{bt}), i.e.,

\be {\delta B \over \delta (\dm)} = 0\, , \label{db0} \ee
 which
 gives the magnetic field minimum value  to be the bottom field $\bff$
when the following condition is satisfied,
\be C/\exp(y)-1 =0\,,
\label{db1} \ee
 which  can give a critical accreted mass,
namely, $\dm = 3.5 {} Log(C) \mcr \simeq  0.5 (\ms) (\mcr/0.2)$,
 after which the bottom field will be preserved
whether or not how much extra mass accreted.  From the statistics
of 24 binary  pulsars
 by van den Heuvel and  Bitzaraki (1995), this critical mass
 is approximately a fraction of $\sim 0.7 M_\odot$.

In Fig.2, the influence of the accretion rate (luminosity) on the
magnetic field is clear that the Eddington  luminosity $L_{38}$
($L_{36}=0.01L_{38}$)
 corresponds to the bottom
field $3\times10^{8}$ G ($3\times10^{7}$ G).

The relation between the X-ray luminosity and
 the magnetic field  was first implied  from
the X-ray spectra of LMXBs by Hasinger and van der Klis (1989),
and they  found that the Z (Atoll) sources with the Eddington
luminosity  ($\sim 1\%$ Eddington luminosity)
possess the strong (weak) magnetic fields.
 However, we stress here that the mass accretion rate cannot be
 exactly inferred from the observations, therefore
 we conclude that the individual Atoll source possesses
a weaker field  than that of Z source  only when
 both sources accretes  the similar masses and the mass
accretion rate of Atoll source is lower than that of Z source.

 Moreover,
the theoretical analysis also implies  the proportional
 correlation between the magnetic field and the X-ray luminosity
 (see, e.g., Psaltis \& Lamb 1998; Compana 2000).
  However, the magnetic field and luminosity relation of
 NS/LMXB  $B \propto L_{x}^{1/2}$  was also  proposed
 by White and Zhang (1997)
when analyzing the kHz QPO data of the ten
 LMXB samples
 under the assumption that the fast spinning  neutron stars are close
 to the spin equilibrium state.
  In reality, for the specific magnetosphere radius $\rrm$ of LMXB,
  if scaled as
$R_M \propto (B^2 /\dot{M})^{2/7} \sim R$, the  bottom  field
should be proportionally related to the accretion rate as $\bff
\propto \dot{M}^{1/2} \propto L_{x}^{1/2}$.

\subsection{The effects of parameter $\xi$}
 If $\xi\ll 1$,  the flow dragging effects on the field lines
 will be low efficiency and the field decay will be slow.
   In other words,
 the small $\xi$ causes decreasing the effective accreted mass
 contributed to the field decay  from Eq.(\ref{bt}).
  The $\xi$ parameter influence can be applicable to the
  binary pulsars PSR 1831-00 and PSR 1718-19, believed to  possess strong
magnetic field $8.7\times10^{10}$ G and $1.28\times10^{12}$,
respectively, after   accreted about half solar mass in the binary
accretion phase  (van den Heuvel \& Bitzaraki 1995). Thus we may
imagine  that the progenitors of both pulsars  experience the
incomplete frozen  plasma flow in their magnetic polar caps, on
account of the unknown instabilities, the
 mechanism of which is still unclear. For these two systems,
  as drawn in Fig.2, if we
set $\xi\sim0.03$ and $\xi\sim0.002$, their polar magnetic field
can decay to the present values from the assumed initial value
$\sim 5\times 10^{12}$ G after accreting half solar mass. While,
the introduction of  $\xi$ parameter helps us to interpret the
sustained  strong  magnetic field NS after accreting $\sim 0.5
\ms$ in systems PSR 1831-00 and PSR 1718-19 (van den Heuvel \&
Bitzaraki 1995), and we find a parameter other than the accreted
mass to influence on the magnetic field decay, which was expected
 by Wijers (1997).
   Moreover, as another example, it
was also pointed out
by Verbunt et al (1990) that NS in 4U 1626-67
has accreted $\sim 0.1 \ms$
without losing its magnetic field $\sim 10^{12}$ G.


\subsection{The NS magnetosphere evolution}

The NS magnetosphere radius  can be defined by the magnetic field
 through the condition  $\rrm/R=(B/B_{f})^{4/7}$,

\be {R_M \over R} = \frac{1}{1 - [C/\exp(y)-1]^2} =
\frac{\exp(2y)}{2C[\exp(y) - C/2]}\,.
\label{rar}
\ee
If $y \ll 1$ or $\dm \ll 3.5 \mcr$, then we have the approximated
correlation between the magnetosphere radius and
 the accreted mass,
\be \rrm = \frac{\rmo}{1 + \frac{4\dm}{7\mb}}\,\,\, or \,\,\,
 {\rrm \ov R} = \frac{1}{{R \ov \rmo} + \frac{8\dm}{7\mcr}}\,. \label{rmdm1} \ee
Furthermore, if $\mb\ll\dm\ll 3.5 \mcr$, then we obtain
 \be \rrm/R
= 0.88({\dm \ov \mcr})^{-1}\,, \label{rmdm2}
\ee
which implies that the magnetosphere radius evolution at this
stage has nothing to do with any initial conditions and even the
X-ray luminosity and only depends on the accreted mass. If the
accreted masses are   similar for the NSs in LMXBs,
 their  magnetosphere radii
   would be homogeneous for both high luminosity Z sources and
low luminosity Atoll sources.
 At the last stage of LMXB,
 if $\dm \sim \mcr$, then we have $\rrm \sim R$.
 The above analytical conclusions are  illustrated in
 Fig.3, where  the magnetosphere evolution  versus the accretion mass
is plotted,  and the accretion rate influence and the
initial field influence are taken into account.
 We find that the magnetosphere radius  is
proportionally related to field strength and inversely related to
the accretion rate at the initial evolution stage, however it has
little correlation to the accretion rate (or luminosity) and the
initial field
 when the the accretion mass exceeds over about $0.01 \ms$.
 In other words, in the late evolution stage,
   LMXB for instance,
    the NS magnetosphere radius is scaled by the NS radius,
    and only depends on the accreted mass.

\subsection{The strong magnetic field estimation in the equator zone}

With the accretion going, the magnetic field lines are dragged by
the accretion flow from the polar cap toward the equator zone, at
where they are  buried inside the crust by the global
compression   of the accreted matter or decayed by
 the possible reconnection
of the field lines in the crust. At the last evolution stage of
LMXB, the global magnetic flux conservation will help us to
estimate the buried equator magnetic field by assuming the initial
magnetic flux to equal the final magnetic flux,
\be 2\pi R^{2}B_{o} \simeq  2\pi R^{2} \bff + 2\pi R H B_{eq}\,.
\ee For the usual condition  $B_{f} \ll B_{o}$, the magnetic field
in the equator zone buried inside the crust can be estimated to be
$B_{eq}\simeq (R/H) B_{o}$, which is much stronger than the
initial field value.  Well, it is remarked that this estimation of
the equatorial magnetic field is based on the assumption of
homogeneously distributed field lines inside the crust.
 On the  detailed magnetic configuration geometry with the accretion,
we refer to the analytical and numerical
calculations by Payne \& Melatos (2004) and Payne (2005), where
they compare the magnetic field line structures  before/after accretion.
 As for the
real NS mass density profile in the crust, the lower the mass
density,  the faster the accretion flow velocity.
 So the more field lines
may be  cumulated in the low mass density layer, where the field
strength would be high but
 lower than  $B_{eqmax}\sim 10^{15}$ Gauss, because
  the fields as large as this would be unstable in the outer crust
and might be able to break the crust at low enough densities
(see, e.g. Thompson \& Duncan 1995).
 The observational evidence to support the
existence of the strong field may occur  in the
Type-I X-ray burst sources,
 which are due to unstable thermonuclear burning of accreted
hydrogen and helium on the NS surface to arise the modulation of
X-rays at the star rotation, and it is believed that the burst is
firstly ignited in the hot spot and then spreads all over star
(see, e.g., Strohmayer \& Bildsten 2003; Cumming 2004).
 Therefore, we associate the hot spot with the existence
 of strong magnetic area.
 Moreover, the accretion-powered X-ray pulsar SAX
J1808.4-3658 (Wijnands \& van der Klis 1998; Wijnands et al 2003)
hints the magnetically channelled accretion, which may also not
exclude the existence of strong magnetic field or the effects of
multiple magnetic moment
 (see, e.g., Psaltis \& Chakrabarty 1999).
 As a plausible application,
recently, Cutler (2002) points out that an internal strong
magnetic  field,
 while  keeping the external dipole magnetic field low,
 may  arise  NS mass quadruple to produce
 that  the  gravitational radiation
braking dominates electromagnetic braking. As a result, NSs  in
various X-ray binaries and  recycled millisecond pulsars could
then be detectable by advanced gravitational wave interferometers.
 In addition,  if the details between  the accretion and polar field
are taken into account, as pointed out by Melatos \& Payne (2005),
 during accretion  the NS magnetic field
  is compressed into a narrow belt at  the magnetic
equator by material spreading equator-ward from the polar cap. In
turn, the compressed field confines the accreted material in a polar
mountain,  which is misaligned with the rotation axis in general,
 to produce the gravitational waves.
At last, we stress that the magnetic field is not totally  decayed
but redistributed on account of the accretion flow, and the polar
field decreases and the equator field increases but buried inside
the crust. The postaccretion diffusion of the suppressed field
back to the NS surface may
 happen in the  recycled pulsars as described by  Young \& Chanmugam (1995).
 Recent observations,
both in the X-ray range of  PSR 1821-14 (see, e.g., Becker et al. 2003),
and in the radio range (see, e.g., Gil et al. 2002),
seem to support the idea of the existence of strong
small-scale magnetic field structures at the NS surface,
 and the further exploration of which has been paid
attention recently by a number of authors (see, e.g., Geppert et
al 2003; Urpin \& Gil 2004).




\section{Discussions and conclusions}

In summary, the main conclusions of model are listed in the
following:
(1) The NS polar magnetic field decays in the binary
accretion phase,
 and its evolution experiences the following processes.
Roughly speaking,  $B\propto \dm^{-1}$ is the initial field
dependent
 if $\dm \sim 10^{-4}\ms$,
 and then    $B\propto \dm^{-1.75}\bff$ is
 independent of the initial field
 after  $\dm \sim 10^{-2}\ms$, until the final stage
 the field B  remaining  stable at the bottom value of about
$B\sim 10^{8}$ G after  $\dm\sim 0.2 \ms$.
(2) The bottom field is the
accretion rate (or luminosity) dependent as $B\propto
\mdot^{1/2}$, which concludes that some  Z (Atoll) sources with
the Eddington  luminosity (1\% Eddinngton luminosity)  may
possess the magnetic fields of about $\sim 3\times 10^{8}$ ($\sim
3\times 10^{7}$) G.
 Therefore,  the existence of the low magnetic field
($\sim 3\times 10^{7}$ G) neutron stars or millisecond pulsars is
suggested, which needs the confirmation from the future
observations.
 (3) The appearance of the bottom field is  nothing
 to do with the initial field strength,  as shown in Fig.2 and Fig.3.
 (4) On the  NS magnetosphere evolution,
   $\rrm \propto \dm^{-1}$ is the initial field dependent
 if $\dm \sim 10^{-4}\ms$ ,
 and then    $\rrm \propto \dm^{-1}$ is
independent of the initial field and the luminosity after $\dm
\sim 10^{-2}\ms$, until the final stage when
 the magnetosphere radius reaches the star radius.
 (5) The equator magnetic field increases with the accretion
 and is stronger
than the initial  field strength, which is buried inside the
crust.
Moreover, as known, the occurrence of thermonuclear X-ray bursts
in LMXBs requires the NSs to have the fields
   $B<10^{10}$ G (see Joss \& Li 1980), so we may
expect the NS in LMXB to accrete $\dm>0.01\ms$ and to enter into
the homogeneous state, while the field B {\it forgets} the initial
field $\bo$ and only {\it remembers} the bottom field $\bff$.
Also, the magnetic field strength distribution of the isolated
non-recycled NSs is from $\sim 10^{10}$ G to
 $\sim 10^{14}$ G (see, e.g., Vranesevic et al 2004; the
highest magnetic field can be extended to  $10^{15}$ G if the
magnetars are included, see Thompson \& Duncan  1995), but most
MSPs locate in a narrow domain of B field distribution, from
$10^{8}$ G to $10^{9}$ G, which seems to support the fact that the
NS final magnetic field  is not  scaled by the initial field $\bo$
but scaled by the bottom  field $\bff$.
 On the characteristic  accreted mass $\mb\sim 10^{-5}$ -- $10^{-3}$ \ms
 for the significant
deduction of the magnetic field, our result is very close
to that obtained by Payne \& Melatos (2004) and Payne (2005), however the
 physical meaning of latter
  is at where the hydrostatic pressure at
the base of
accretion column overcomes the magnetic tension and the matter
spreads over the stellar surface to drag the polar  field lines
toward the equator.
 In our model, $\mb$ represents the fraction of the crust mass
scaled by the ratio of the stellar radius to the initial
magnetosphere radius, then the relation  of both characteristic
masses is still unclear.  Finally, although our model has given the
observational consistent results, it is a very simple and crude
theoretical framework to account  for the accretion induced polar
field decay. We think that the   basic physical  idea is clear and
has been proposed by a couple of people (see, e.g., van den Heuvel
and Bitzaraki 1995; Burderi et al 1996; Burderi \& D'Amico 1997)
that the transformation from the systematic channeled accretion to
the spherical accretion yields the power to decrease the NS  polar
field strength. However, many physical   details have  been
neglected in the model, such as, the detail of accretion flow and
its reaction with the field lines, the plasma instability in the NS
melted surface,
 the details of how the  magnetic field is buried in the
equatorial zone, the Ohmic dissipation, as well as the
Hall-drift effects,  etc., therefore the
considerations  of these will construct our future
research explorations.

\section*{Acknowledgements}
It is a  pleasure to thank K.S. Cheng, U. Geppert, G. Hasinger,
J.L. Han, D. Lai,  Q.H. Luo, R.N. Manchester,
 D.B. Melrose,  G.J. Qiao,  R.X. Xu, X.J. Wu
 for helpful discussions. This research has been supported by the innovative
project of CAS of China.
  The author expresses  the  sincere thanks to the critical comments
  from the anonymous referee  that greatly improved the quality of the paper.

%

\begin{figure}
\hspace{3mm}
\includegraphics[height=8cm]{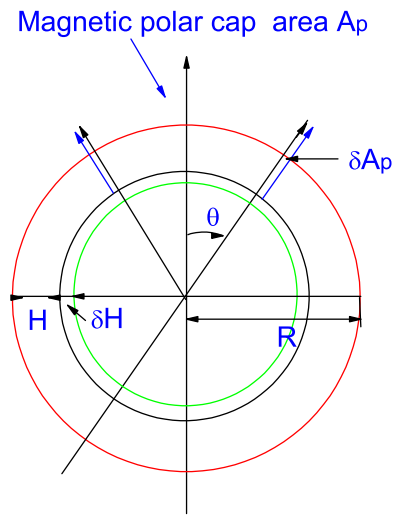}
\caption{The cross section of the accreted neutron star. $\theta$ is
the magnetic polar cap  angle of neutron star with  radius R. H is
the thickness of the crust and  $\delta H$ is the expanded immersed
depth. $A_{p}$ and $\delta A_{p}$ are the areas of magnetic polar
zone and its variation respectively.}
\end{figure}

\noindent \hspace{-2cm}
\begin{figure}
 \includegraphics[width=5.5cm, angle=270]{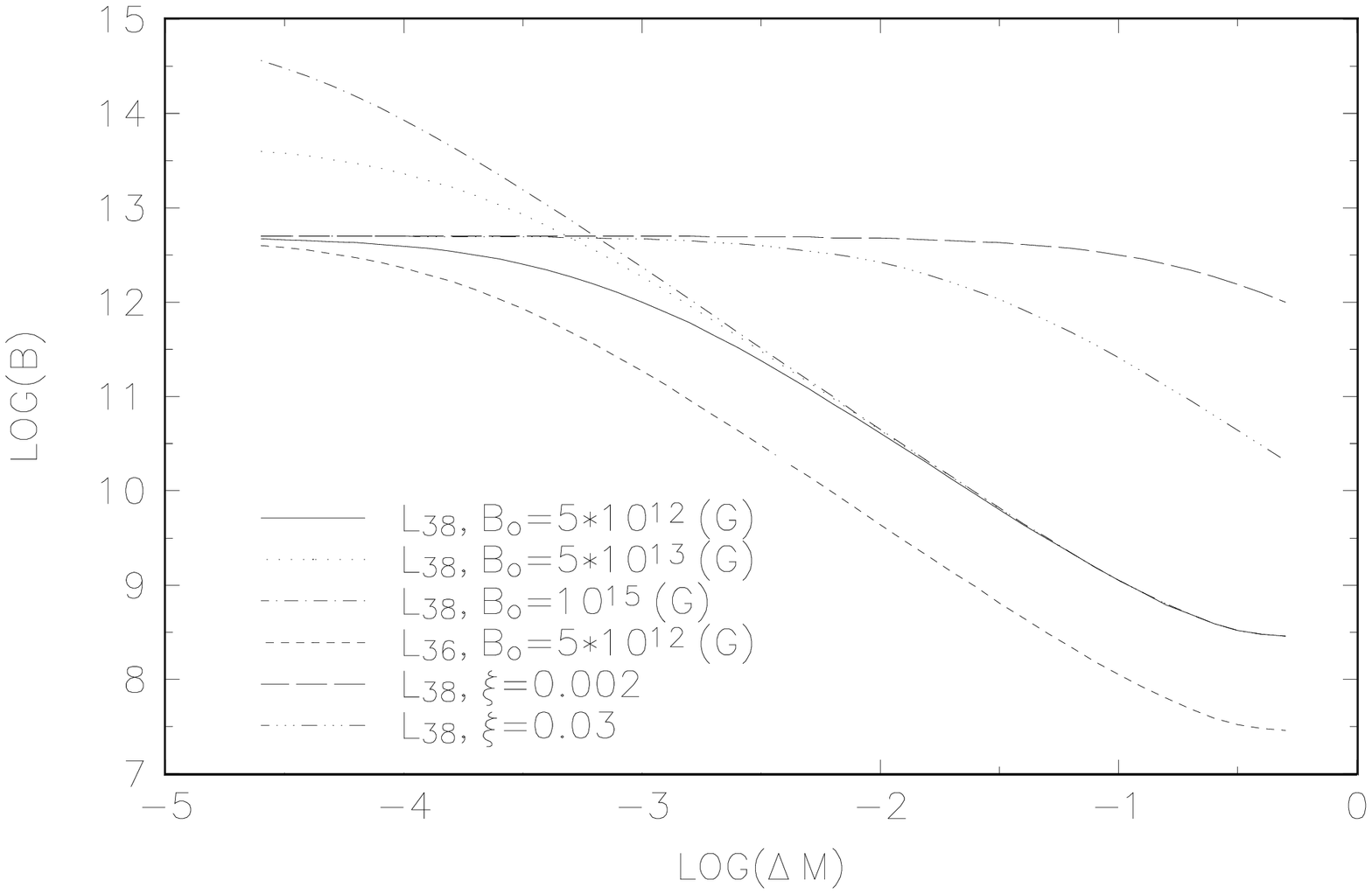}
\vskip 1.83cm
  \caption{ The magnetic
field versus accreted mass diagram with various parameter
conditions, the Eddington rate $L_{38}$ and $L_{36}=0.01L_{38}$,
 and the initial field strengths from $\bo  =
10^{12}$ G to $\bo = 10^{15}$ G, which are indicated in the figure.}
\end{figure}

\noindent
\begin{figure}
\includegraphics[width=5.5cm,angle=270]{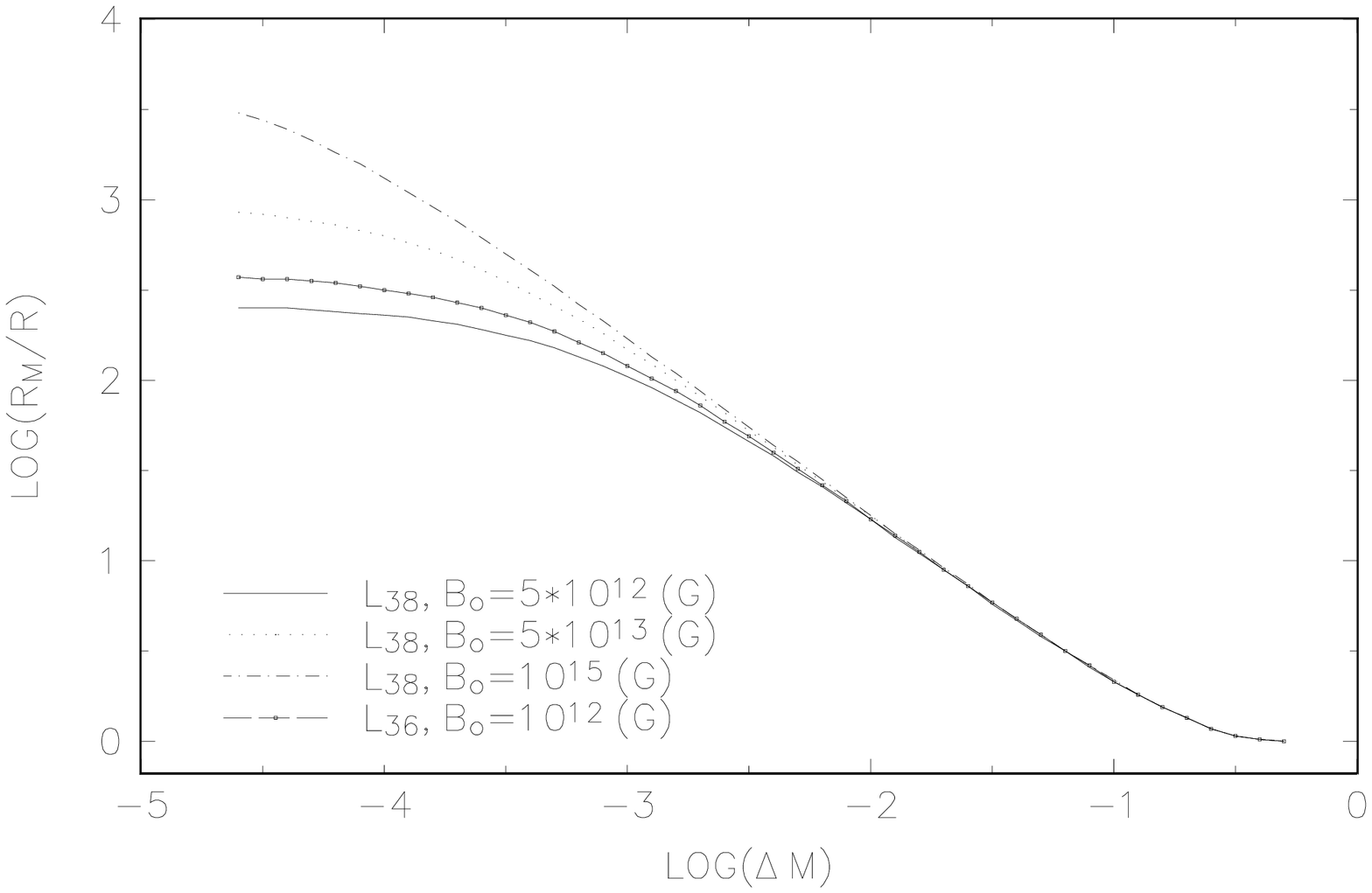}
\vskip 1.83cm
 \caption{ The magnetosphere
radius versus accreted mass diagram with various parameter
conditions, the Eddington rate $L_{38}$ and $L_{36}=0.01L_{36}$,
and the initial field strengths from $B_0 = 10^{12}$ G to $B_0 =
10^{15}$ G, which are indicated in the figure.}
\end{figure}

\end{document}